\documentclass[a4paper,11pt]{article}
\pdfoutput=1 

\usepackage{jcappub} 

\usepackage[T1]{fontenc} 
\usepackage{booktabs} 
\usepackage{ dsfont }
\usepackage{float}

\usepackage{calrsfs}
\DeclareMathAlphabet{\pazocal}{OMS}{zplm}{m}{n}
\usepackage{xcolor}
\usepackage{pict2e}
\usepackage[percent]{overpic}
\usepackage{ upgreek }
\usepackage{xfrac}
\usepackage{abraces}
\usepackage{stmaryrd}
\usepackage{amsmath,empheq}
\usepackage{trimclip}
\usepackage{braket}
\usepackage{subcaption}
\usepackage{stfloats} 
\usepackage{units}

\makeatletter
\DeclareRobustCommand{\shortto}{%
  \mathrel{\mathpalette\short@to\relax}%
}

\newcommand{\short@to}[2]{%
  \mkern2mu
  \clipbox{{.5\width} 0 0 0}{$\m@th#1\vphantom{+}{\shortrightarrow}$}%
  }
\makeatother

\title{Superdense beaming of axion dark matter in the vicinity of the light cylinder of pulsars}

\author[a]{Javier De Miguel} 
\author[a]{and Chiko Otani}

\affiliation[a]{The Institute of Physical and Chemical Research (RIKEN), Center for Advanced Photonics, 519-1399 Aramaki-Aoba, Aoba-ku, Sendai, Miyagi 980-0845, Japan}

\emailAdd{javier.miguelhernandez@riken.jp}

\abstract{In this article we treat the non-adiabatic photon-to-axion resonant conversion of curvature radiation, synchrotron emission and inverse Compton scattering dominating the spectral density function of pulsars. First, we introduce emission models and benchmark observational data. We adopt a state-of-the-art density profile that relieves tension with the quantum electrodynamics vacuum polarization effect in highly magnetic stars, leading to efficient mixing. Then, we estimate the dark matter flux induced by photon-axion oscillation across the light cylinder of the neutron star. We find that pulsars might produce axion overdensities many orders of magnitude over the occupation number of dark matter in the Galactic halo within a broad parameter space. We point out possible new methods for axion detection derived from these results and other future lines of work.}


\begin{document}
\maketitle
\flushbottom 

\section{Introduction}
\label{sec:Intro}
Axions \cite{PhysRevLett.40.223,PhysRevLett.40.279} are hypothetical pseudo-scalar bosons theorized as a consequence of the dynamic solution to the problem of charge and parity (CP) in the strong interaction \cite{PhysRevLett.38.1440}. Furthermore, axion is a well-grounded candidate for cold dark matter (cDM) \cite{ABBOTT1983133,DINE1983137,PRESKILL1983127,PhysRevD.98.030001}. Axion and axion-like particles (ALPs) mix with photons by action of an external magnetic field. The axion-photon coupling Lagrangian density is 

\begin{equation}
\pazocal{L} \supset -\frac{1}{4} F_{\mu \nu} F^{\mu \nu}
\!+\frac{1}{2} \partial_{\mu} \phi\, \partial^{\mu} \!\phi\, 
-\frac{1}{2} \mathrm{m^2_{\phi}} \,\phi^2 
-\frac{\mathrm{g}_{\phi\gamma}}{4} F_{\mu \nu} \tilde{F}^{\mu \nu}\phi -j^{\mu} A_{\mu}
\,,
\label{Eq.1.1}
\end{equation}
being $F^{\mu \nu}$ the field strength tensor and $\tilde{F}$ its dual, $\phi$ the axion field, $\mathrm{m_{\phi}}$ the pseudo-scalar mass and $\mathord{\mathrm{g}}_{\phi \gamma}$ the axion-photon coupling constant, $j$ is density of current and the standard photon field is $A^{\mu}$. Classically, the interaction term in Eq. \!\ref{Eq.1.1} simplifies to $\mathord{\mathrm{g}}_{\phi\gamma}  \phi \, \mathord{\mathrm{E}} \cdot \mathord{\mathrm{B}}$, with $\mathbf{\mathrm{E}}$ the photon field and $\mathbf{\mathrm{B}}$ the external magnetic field.

The quantum chromodynamics (QCD) axion presents a light mass which scales inversely to a typical energy scale, $f_{\!\phi}$. This parameter determines the axion-photon coupling strength. Differently, ALPs present mass-coupling independent parameters, creating a large parameter space which has been only superficially explored — cf. \cite{PhysRevD.98.030001}. \newline

Neutron stars (NSs) are evolved compact objects with extreme surface magnetic fields \cite{Petri:2016tqe}. The idea of axion-to-photon conversion in the magnetized plasma of NSs was pioneered decades ago \cite{Iwamoto:1984ir, Raffelt:1986np, Morris:Phys.Rev.D34, Raffelt:1987im, Raffelt:1987np, Yoshimura:1987ma}. This concept has been revisited by numerous authors over the years — e.g., see \cite{Lai:2006af, Witte:2021arp} —. More recently, the radio signals produced by infalling axions through axion-to-photon conversion in magnetospheres has been studied \cite{Huang:2018lxq, Hook:2018iia, Safdi:2018oeu, Leroy:2019ghm, Battye:2019aco}; also for collisions in \cite{Iwazaki:2014wka, Bai:2017feq, Buckley:2020fmh}. Complementarily, in \cite{Perna:2012wn, Zhuravlev:2021fvm} authors employ the formalism in \cite{Lai:2006af} to estimate the signatures in the thermal spectra of NSs caused by photon-to-axion oscillation. Parallel non-thermal axion-photon conversion mechanisms in the environment of NSs have been proposed lately \cite{Garbrecht:2018akc,Prabhu:2021zve}. \newline

In the present work we investigate the non-adiabatic resonant conversion of curvature radiation, synchrotron emission and inverse Compton scattering photons dominating the spectral density function of isolated neutron stars. We apply this concept to spectral functions from astronomical observations in order to estimate the overdensity of axion dark matter that might be beamed from somewhere around the light cylinder of rotation-powered pulsars and B-powered magnetars.

The manuscript is structured as follows. First, we briefly explain the fundamentals of neutron stars relevant for this work in section \ref{sec:NS}. The axion-photon mixing equations are introduced in section \ref{sec:Prob} in order to obtain an analytical expression which allows us to estimate the photon-to-axion conversion probability in the surroundings of the star, and we apply this formalism to realistic spectral density functions. The results are presented in section \ref{sec:Results}, and discussed in section \ref{sec:Discussion}.

\section{Rotation-powered pulsars and magnetars}
\label{sec:NS}
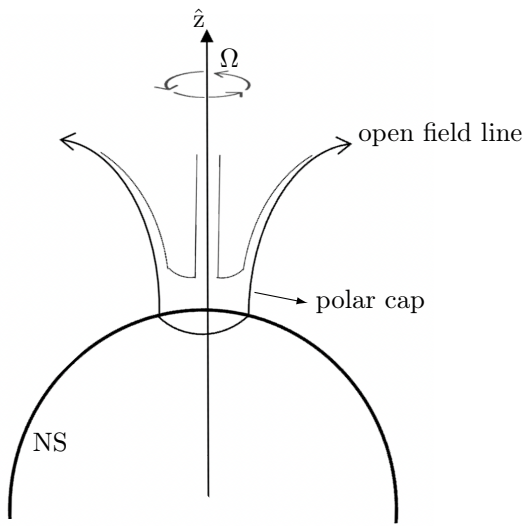
\begin{figure}[ht!]
		\includegraphics[width=0.4\textwidth]{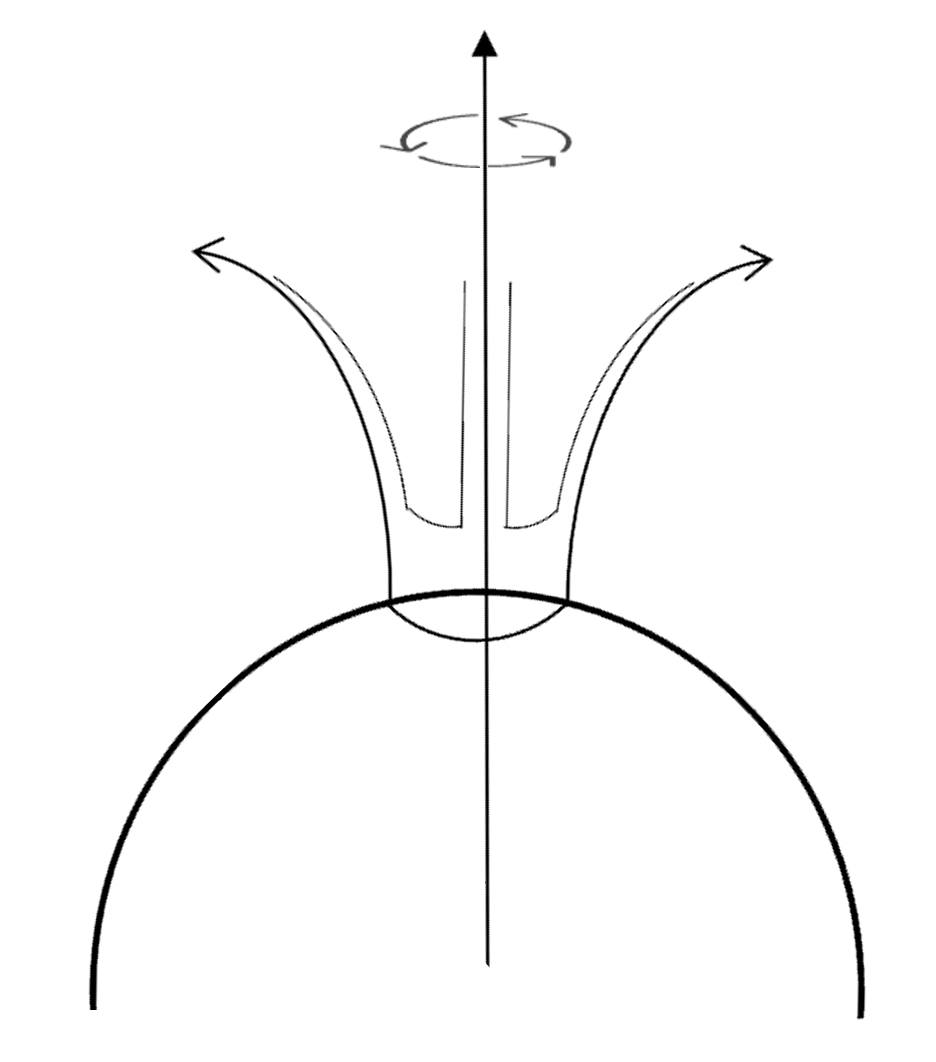}
		\centering 
		\begin{small}
		\put(-150,35){NS}
		\put(-80,180){$\Omega$}
		\put(-90,195){$\mathrm{\hat{z}}$}
		\put(-44,89){polar cap}
        \put(-67,95){\vector(1,-0.2){20}}
        \put(-28,152){open field line}
        \end{small}
		\caption{Non-scale representation of a neutron star (NS). In the `aligned rotator' approximation, the magnetic axis and the rotation axis are superimposed.}
		\label{fig_1}
	\end{figure}

 Pulsating stars (pulsars) are rotating neutron stars with regular periods. `Canonical' neutron stars would present a radius about $R\!\sim$10 km, a mass of the order of $M\!\!\sim$1.4 $\!\mathrm{M}_{\odot}$. In a pseudo-aligned rotator, the field becomes stationary, or $\mathord{\mathrm{B_z}}\!\simeq\!B_{\!_0}\!\left(\frac{R}{r}\right)^{\!3}$, and the dielectric tensor does not vary significantly during a period. Here, $B_{\!_0}$ is the field strength at the surface of the NS, of the order of $10^{8-10}$ T, or $\gtrsim\!\!10^{10-11}$ T in the case of magnetars \cite{Duncan:1992hi,Kaspi:2017fwg,Turolla:2015mwa}; $r$ is the radial coordinate. The angular velocity is $\Omega\!=\!2\pi/P$, being $P$ a rotation period from tens to milli- seconds. The surface temperature is around $10^6$ K. Their thin atmosphere is composed of a `cold' plasma — mainly H and He — \cite{2012MNRAS.423.1556S}. A neutron star is illustrated in Fig. $\!$\ref{fig_1}, where some aspects that are relevant to this article are highlighted.
 
Even if discovered half a century ago \cite{Hewish:1968bj}, pulsar emission mechanisms remain full of incertitude \cite{2009MNRAS.397..103S}. The standard picture is based on the creation of pair cascades and particle migration trough magnetic forces, accelerating the charges to relativistic velocities. Radiative processes are dominated by curvature radiation, synchrotron emission and inverse Compton scattering. However, acceleration and emission regions, or the specific altitude — sometimes referred to as $h_{0}$ \cite{Harding:1998ma} — at which the pair cascade and charge acceleration start, is a prominent point of discrepancy. The classic picture considers radio emission from pulsars is originated at 
`low altitudes' over the polar cap \cite{1975ApJ...196...51R, 1979ApJ...231..854A, 1997A&A...322..846K}. The high-energy emission is regularly associated to a wider interval of higher altitudes. Nevertheless, the precise concept of `low altitude' and `high altitude' is diffuse, and different models suggest that high-energy emission originates somewhere inside the light cylinder (LC), around the LC, etc. — cf. \cite{Petri:2016tqe,2009MNRAS.397..103S} and references therein.  

The light cylinder is filled with plasma at a corotation density. Goldreich \& Julian (GJ) calculated the minimal corotational charge density — $n_{_\mathrm{{GJ}}}$ — in a neutron star under a number of idealizations, including a perfect conductivity and homogeneity \cite{1969ApJ...157..869G}. Frequently, modern approaches are over-dense compared to the GJ model, and a pair multiplicity factor $\kappa\!=\!n_{e}/n_{_\mathrm{{GJ}}}$ arises with typical values $10^{4-7}$ \cite{2000ApJ...532.1150Z, Jessner 2001, Eilek:2016hms, Guepin:2019fjb}. In that manner, the charge density in our model of star reads

\begin{equation}
n_{e}=\frac{2\,\varepsilon_{_0}\Omega B(r)}{e} \, \kappa \,,
\label{Eq.4.1}
\end{equation}

while the characteristic frequency of the plasma is
\begin{equation}
\omega_p=\left( \frac{e^2 n_{e}}{m_e\varepsilon_{_0}} \right)^{\!\!\!1/2}.
\label{Eq.4.2}
\end{equation}

\begin{figure}[ht!]
		\includegraphics[width=0.75\textwidth]{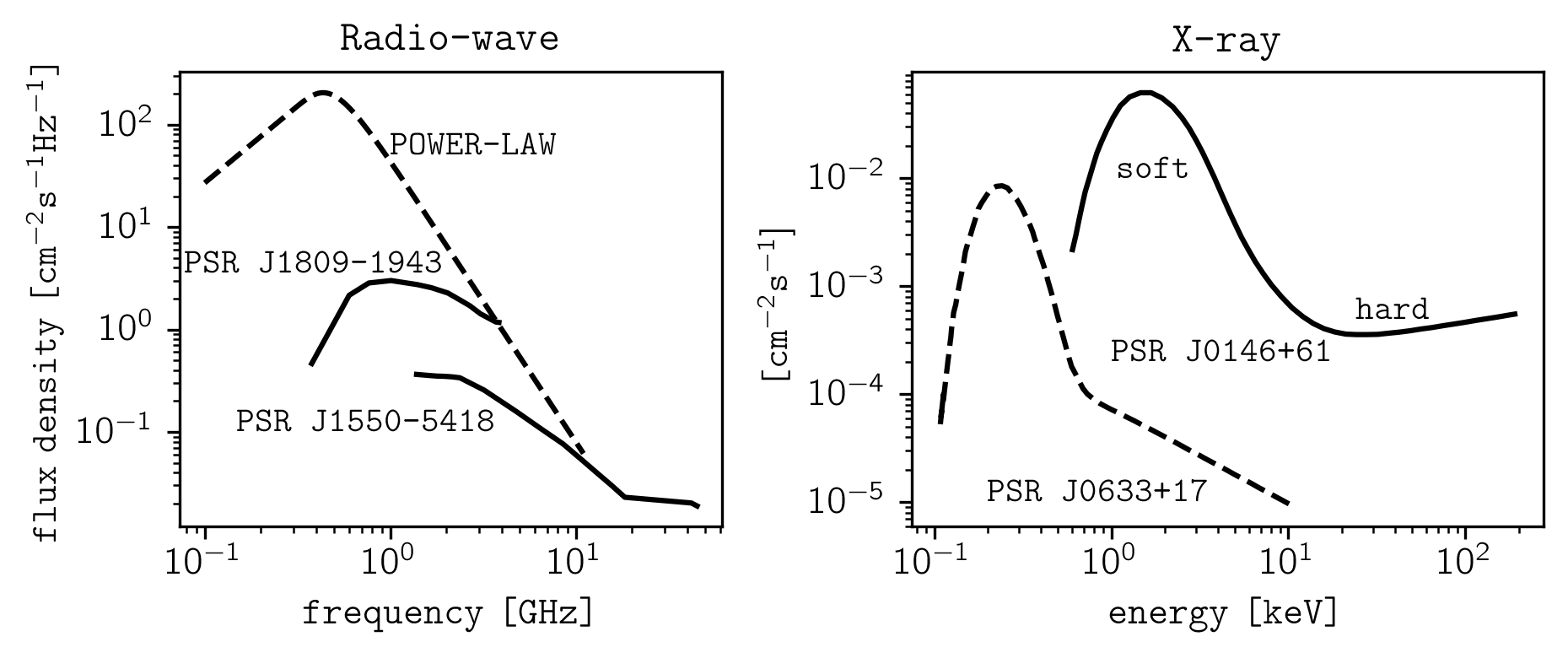}
		\centering 
		\caption{Spectral flux density received at the Earth position for rotation-powered pulsars — dashed line — and magnetars — solid line — in the radio-wave — left —, or `soft' and `hard' X-ray high-energy range — right. Typical units are used for each case. Distances are $\pazocal{O}(\nicefrac{1}{2})$ kpc for a rotation-powered pulsar modeled by the power-law Eq. $\!$\ref{Eq.2.1}, $\mathrm{r}\!\sim$6 kpc for PSR J1550--5418, $\mathrm{r}\!\sim$3.5 kpc for PSR J1809--1943, $\mathrm{r}\!\sim$0.2 kpc for Geminga pulsar — PSR J0633+17 — and $\mathrm{r}\!\sim$2.7 kpc for magnetar PSR J0146+61. PSR J1809--1943 is included for completeness, showing a very steep slope at low-frequency.}
		\label{fig_2}
	\end{figure}

Observational campaigns are a reliable alternative in order to provide luminosity curves and spectral densities in the absence of a well-established theoretical model. In Fig. $\!$\ref{fig_2} we present the reference spectral density functions for this article. The radio-frequency (RF) spectra is modeled using Eq. $\!$\ref{Eq.2.1} \cite{Torne:2015rha,Maron:2000wn,Jankowski:2017yje,Bates:2013ear,Murphy:2017ech}, or data from \cite{Cerutti:2013kza, Kargaltsev:2015pma, Camilo:2008ns} for magnetar PSR J1550--5418 — an object of interest as its rotation and magnetic axis appear to be lined up — and magnetar PSR J1809--1943 \cite{Maan:2019ext}. The X-ray spectra, adapted from \cite{Rea:2007xd, Mereghetti:2015asa}, corresponds with magnetar PSR J0146+61. The Geminga — rotation-powered — pulsar PSR J0633+17 spectral function is adapted from \cite{Mereghetti:2015asa, Jackson:2005ms}. 

The polarization state of synchrotron emission is governed by the magnetic vector. Even though that the polarization of the photon beam from a neutron star can be affected by Faraday rotation and other spurious effects through line-of-sight to observer, recent research could shed some light about its properties. Interestingly, the persistent signal received from neutron stars presents a strong linear polarization in the direction of the rotation axis, with Stokes V$\simeq\!0$ \cite{2009MNRAS.397..103S, Forot:2008ud, Petri:2016tqe, Petri:2005ys}. This parallel component of the photon field refers to the ordinary mode — O-mode — that we confine our concern throughout this work.

\section{Non-adiabatic photon-axion resonant oscillation in neutron stars}
\label{sec:Prob} 
In this section we focus on O-mode waves propagating in the `weak dispersion' limit with a refractive index close to unity, or equivalently $\omega \!\!\sim \mid \!\!\mathrm{k}\! \!\mid$. In a strongly magnetized `cold' plasma, the conductivity is non-zero only along the direction of the magnetic
field \cite{1993ppm..book.....B} — appendix \ref{C} —. The photon-axion oscillation takes place `on-the-spot', or in a narrow projection along $r_{\!c}$ and hence the gravitational gradient is negligible. In that manner, the wavelength becomes stationary and arbitrarily shifted to the red. The mixing equations are derived from Eq. $\!$\ref{Eq.3.2b}  \cite{Raffelt:1987im}

\begin{equation}
i \frac{d}{dz} \begin{pmatrix}\phi \\  \mathord{\mathrm{E}}_{_{\parallel}} \end{pmatrix} 
= \begin{pmatrix} \omega+\Delta_{\phi}+\Delta_{\mathrm{{vac}}_{\parallel}} & \Delta_{\mathrm{g}_{\phi\gamma}} \\ \Delta_{\mathrm{g}_{\phi\gamma}} & \omega- \Delta_p -\Delta_{_{\parallel}} 
+ \Delta_{\mathrm{{vac}}_{\!\perp}}
\end{pmatrix} \begin{pmatrix} \phi \\  \mathord{\mathrm{E}}_{_{\parallel}} \end{pmatrix}  \,,
\label{Eq.3.1}
\end{equation}
 where we introduced $\Delta_p=\frac{\omega^2_p}{2\omega} \mathrm{sin}^2 \theta$, $\Delta_{_{\parallel}}=\frac{1}{2}\omega (\mathrm{n_{_{_{\parallel}}}}-1)$, $\Delta_{\phi}=-\frac{m^2_{\phi}}{2\omega}$ and $\Delta_{\mathrm{g}_{\phi\gamma}}\!\!=2 \mathrm{g}_{\phi\gamma} \mathrm{sin}^2 \theta$. We also define $\Delta k= 2[(\Delta_{\phi}-\Delta_p-\Delta_{_{\parallel}})^2/4+\Delta^2_{\mathrm{g}_{\phi\gamma}}]^{1/2}$ and the mixing angle $ \mathrm{tan}(2\theta_m)=2\Delta_{\mathrm{g}_{\phi\gamma}}/(\Delta_{\phi}-\Delta_p-\Delta_{_{\parallel}})$. The term $\Delta_{_{\parallel}}$ vanishes in the weak dispersion limit as the refractive index is $\mathrm{n_{_{_{\parallel}}}}\!\!\simeq k/\omega \!= \!1+ \Delta_{_{\parallel}}/\omega$. Quantum electrodynamics (QED) modifications to Maxwell's equations due to vacuum polarization dissipate in dense sectors, where vacuum effects are negligible compared to plasma effects and, therefore, the vacuum refractive index — $\Delta_{\mathrm{{vac}}}$ — vanishes from the dispersion relations of Eq.$\,$\ref{Eq.3.1} in certain circumstances — see appendix \ref{E}.

Landau-Zener-Stueckelberg-Majorana formalism  \cite{Landau:1932,1932RSPSA.137..696Z,Stueckelberg,1932NCim....9...43M} provides an analytic solution to the mixing equations of a two-state system with an energy gap between the states which depends linearly on time.\footnote{Landau-Zener(-Stueckelberg -Majorana) formula allows one to solve a Schrödinger-like equation in the form $i\frac{\partial}{\partial z} \uppsi_{\!\mu} =\pazocal{H}(z) \uppsi_{\!\mu}$, where $\pazocal{H}(z)$ is a lineal Hamiltonian and $\uppsi_{\!\mu}$ is the state tensor. Perturbative analysis provide the oscillation amplitude $\braket{\phi|\pazocal{H}_{\mathord{\mathrm{int}}}|\mathord{\mathrm{E}}_{_{\parallel}}}$, being $\pazocal{H}_{\mathord{\mathrm{int}}}$ the density of perturbation of the interaction term in Eq. \ref{Eq.1.1} \cite{Raffelt:1987im}. In \cite{Battye:2019aco} authors confirm that both Perturbation theory and Landau-Zener formalism — employed in \cite{Lai:2006af} —  yield an equivalent result, as expected.} The transition probability between the states — $\upzeta,\upeta$ — reads $\mathord{\mathrm{P}}\!\!_{_{\upeta \leftrightarrow  \upzeta}}\!\!=\!1-e^{-2\pi\Gamma}$, with $\Gamma$ the adiabaticity parameter of the system. Now, in order to write-down a compact analytic expression from Eq. $\!$\ref{Eq.3.1}, we consider pure non-adiabatic conversion so that the characteristic parameter results $\Gamma\!\!<\!<\!1$ at resonance \cite{Lai:2006af, Battye:2019aco, Perna:2012wn, Zhuravlev:2021fvm}. A second-order series truncation yields $\mathord{\mathrm{P}}\!\!_{\gamma\shortto \phi}\!\simeq\! 2\pi\!\!\mid\!\!\Gamma\!\!\mid$, with $\Gamma\!=\Delta\!^2k/2/\frac{\partial}{\partial z}\Delta_p$. For outgoing beams \cite{Giraud} we get $\frac{\partial}{\partial z} \omega_p\!\sim \mathrm{m_{\phi}}/r_{\!c}$ and hence $\frac{\partial}{\partial z}\Delta_p\!=-\frac{1}{\omega}\omega_p \frac{\partial}{\partial z} \omega_p \!\sim \mathrm{m^2_{\phi}}\omega^{-1}r_{\!c}^{-1}$ — see appendix \ref{D} —. It is straightforward to obtain

\begin{equation}
\mathord{\mathrm{P}}\!_{\gamma \shortto \phi}= \frac{\mathrm{g}^2_{\phi\gamma} B^{2}(z_c) \,\omega \,z_c}{\mathrm{m^2_{\phi}}}  \;\; + \pazocal{O}(\Gamma^2)\,.
\label{Eq.3.2}
\end{equation}

The expression in Eq. $\!$\ref{Eq.3.2} unveils interesting physics. The conversion probability scales with the frequency of the photon from the NS, $\omega$. Contrarily, the axion mass range for which resonant conversion takes place is set by the density profile through the magnetosphere, since $\omega_p\!=\!\mathrm{m_{\phi}}$ at resonance. That is conceptually similar to the case of haloscopes, where the scanning frequency is only given by the resonant frequency of the cavity \cite{PhysRevLett.51.1415}; or light-shining-through-walls (LSW) experiments, where the mass of the axion emerging in the axion-photon vertex is determined by the resonant frequency of the cavity independently of the nominal frequency of the laser pump \cite{VanBibber:1987rq, OSQAR, Ehret:2010mh}. Consequently, the conversion probability depends only on two parameters for the stated model of star and axion, i.e., $z_c$ and $\omega$. Moreover, given the $\omega$ factor in Eq.$\!$ \ref{Eq.3.2}, non-adiabatic conversion is inconsistent when entering the $\gamma$-ray energy frame. That is the reason why a $\gamma$-ray analysis is out of the scope of the article. Finally, the production of axions with lower coupling factors is enhanced at high frequency. 

In order to render feasible an analytic approach, we consider a stationary magnetic field and a strongly magnetized plasma \cite{Lai:2006af, Witte:2021arp, Hook:2018iia, Safdi:2018oeu, Leroy:2019ghm, Battye:2019aco, Perna:2012wn, Zhuravlev:2021fvm}. On the other hand, QED corrections suppress the cross-section by orders of magnitude  when vacuum polarization effects are significant \cite{Raffelt:1987im}. Consequently, our non-adiabatic resonant model is ruled out in several sectors. First, O-modes are evanescent below the cut-off frequency of the plasma \cite{Kunzl:1998fr, Melrose:2020mhu}, when $\omega_p\!>\!\omega$. Second, adiabatic regions with $\Gamma \!\!\gtrsim \!\!1$ appear near the stellar surface for X-rays. Third, for sectors with $\omega\!\gtrsim\!\omega_{\!_B}$, with $\omega_{\!B_{\!c}}\!\!=\!e B(z_c)/m_e$, standard simplifications in the dielectric tensor do not hold — see appendix \ref{C} —. Finally, the parameter space at which QED corrections govern the dispersion relations in Eq.$\,$\ref{Eq.3.1}, or equivalently $\omega^2_p\!<\!Q_{\mathrm{QED}}$ — consult appendix \ref{E}.

The time-scale of the photon-axion system at resonance, $\tau_{\!s}\!\sim \mid\!\! \omega-\omega_p\!\!\mid^{-1}$, is typically much shorter than the pulsation time-scale of the star — from the tens ns up to few ms in the radio-frequency regime \cite{Hankins:2016bbl}; significantly larger in the case of X-rays \cite{Kaspi:2017fwg} —. Then,  $\tau_{\!s}\!\!<\!<\!t_{\mathrm{pulse}}$ and the conversion takes place in a pseudo-stationary regime instead of transitory, where the model could loss generality.

\begin{figure}[ht!]
		\includegraphics[width=0.57\textwidth]{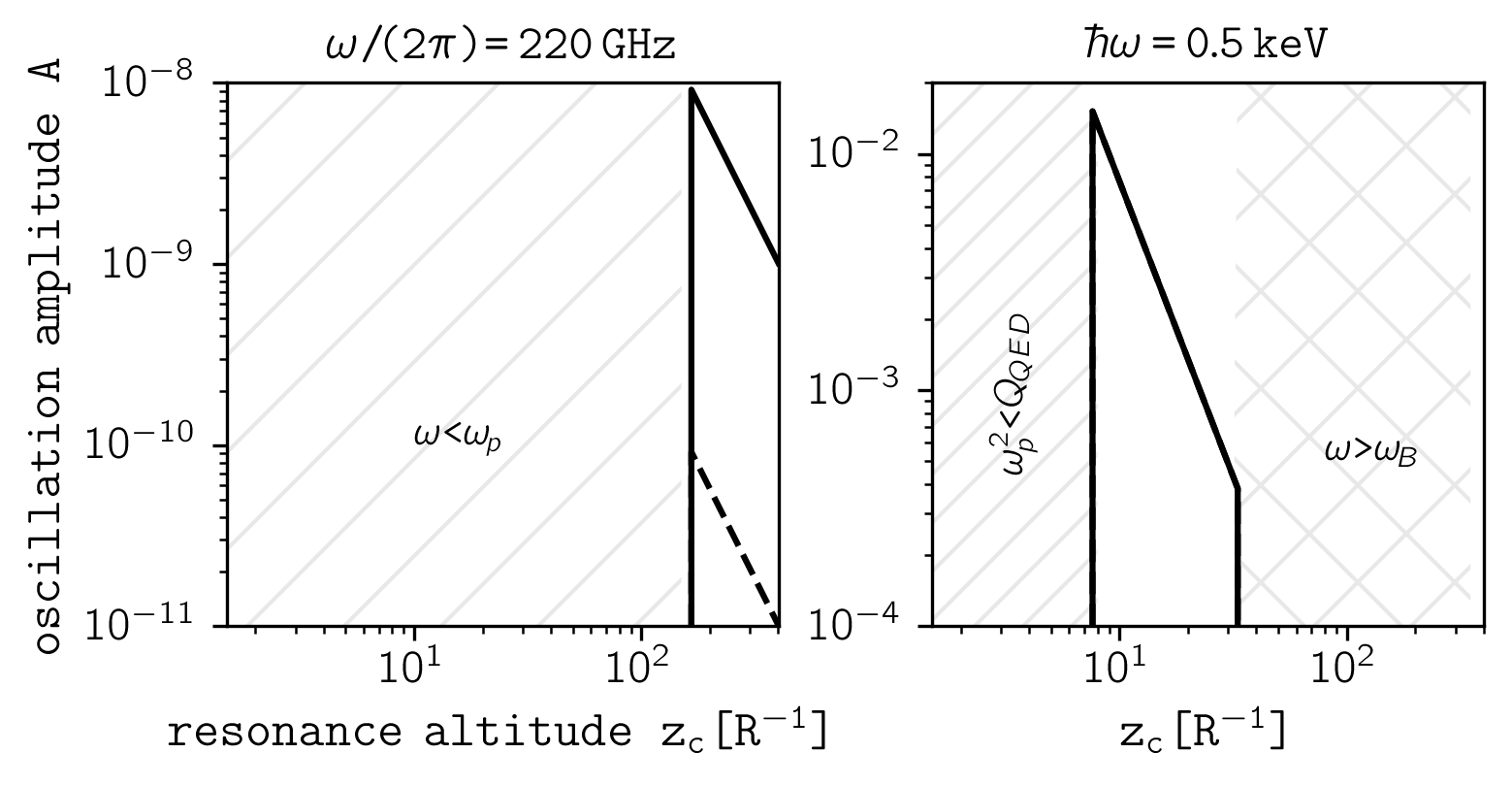}
		\centering 
		\caption{Photon-axion oscillation amplitude — $\mathrm{A}$ — over distance from the stellar surface — $z_c$ — for different threshold frequencies. Rotation-powered pulsars — dashed line — and magnetars — solid line — are included. Parameters are $P$=1 s, $B_{\!_0}$ is $5\!\times\!10^8$ T for a rotation-powered pulsar and $5\!\times\!10^{10}$ T for the magnetar. The pair multiplicity factor is $\kappa\!\!\sim\!\!10^{6}$. We adopt KSVZ-like coupling strength \cite{PhysRevLett.43.103,SHIFMAN1980493} and canonical dimensions. The non-adiabatic resonant model is ruled out in the following sectors: (i) regions where O-modes become evanescent, as $\omega_p\!>\!\omega$; (ii) adiabatic sector with $\Gamma\!\gtrsim\!1$; (iii) `weak' field zones, where $\omega\!\gtrsim\!\omega_{\!_B}$; (iv) parameter space at which QED vacuum effects govern the dispersion relations, or $\omega^2_p\!<\!Q_{\mathrm{QED}}$.}
		\label{fig_3} 
	\end{figure}

Oscillation amplitudes — $\mathrm{A}_{\gamma \shortto \phi}\!= \!\Gamma^{\nicefrac{1}{2}}$ — are represented in Fig. $\!\!$\ref{fig_3} for benchmark parameter sets. 

\section{Results}
\label{sec:Results}
We have established in previous sections that, once the axion and NS models are assigned, the non-adiabatic resonant conversion probability in Eq.$\!$ \ref{Eq.3.2} depends only on the frequency of the photon emitted by the NS — $\omega$ — and the distance from its surface — $z_c$ —. The latter parameter determines the charge density via Eqs. \ref{Eq.4.1} and \ref{Eq.4.2} —. The resulting spectral flux density of the axion beam produced in the axion-to-photon conversion reads

\begin{equation}
\Phi_{\!\phi}= \Phi_{\!\gamma} [\mathrm{cm^{-2}\, s^{-1} \, Hz^{-1}}] \times \mathord{\mathrm{P}}_{\!\!\gamma \shortto \phi}        \,.
\label{Eq.4.4}
\end{equation}
Finally, we define the dark matter (DM) spectral flux density

\begin{equation}
\Phi_{\!\mathrm{_{DM}}}= \rho_{_{\!\oplus}} [\mathrm{eV \,\! cm^{-3}}] \times v_{\phi}\, [\mathrm{cm\,s^{-1}}] \frac{\,\mathrm{eV}\,}{\mathrm{m_{\phi}}} \frac{\,\mathrm{Hz}\,}{\Delta{\nu}_{\!\phi}} \, ,   
\label{Eq.4.5}
\end{equation}

where $\rho_{_{\!\oplus}}$ is the occupation number of dark matter around the Solar System position, about $\rho_{_{\!\oplus}} \!\!\sim 300$ MeV$\!\,$cm$^{-3}$, and $\Delta{\nu}_{\!\phi}\!\sim \!10^{-6}\nu$ is the line-width of the axion signal for the virial velocity $v_{\phi}\!\sim\!10^{-3}c$.

\begin{figure*}[ht!]
\begin{subfigure}{0.52\linewidth} \hspace{-12pt}
    \includegraphics[width=\linewidth, height=0.52\linewidth]{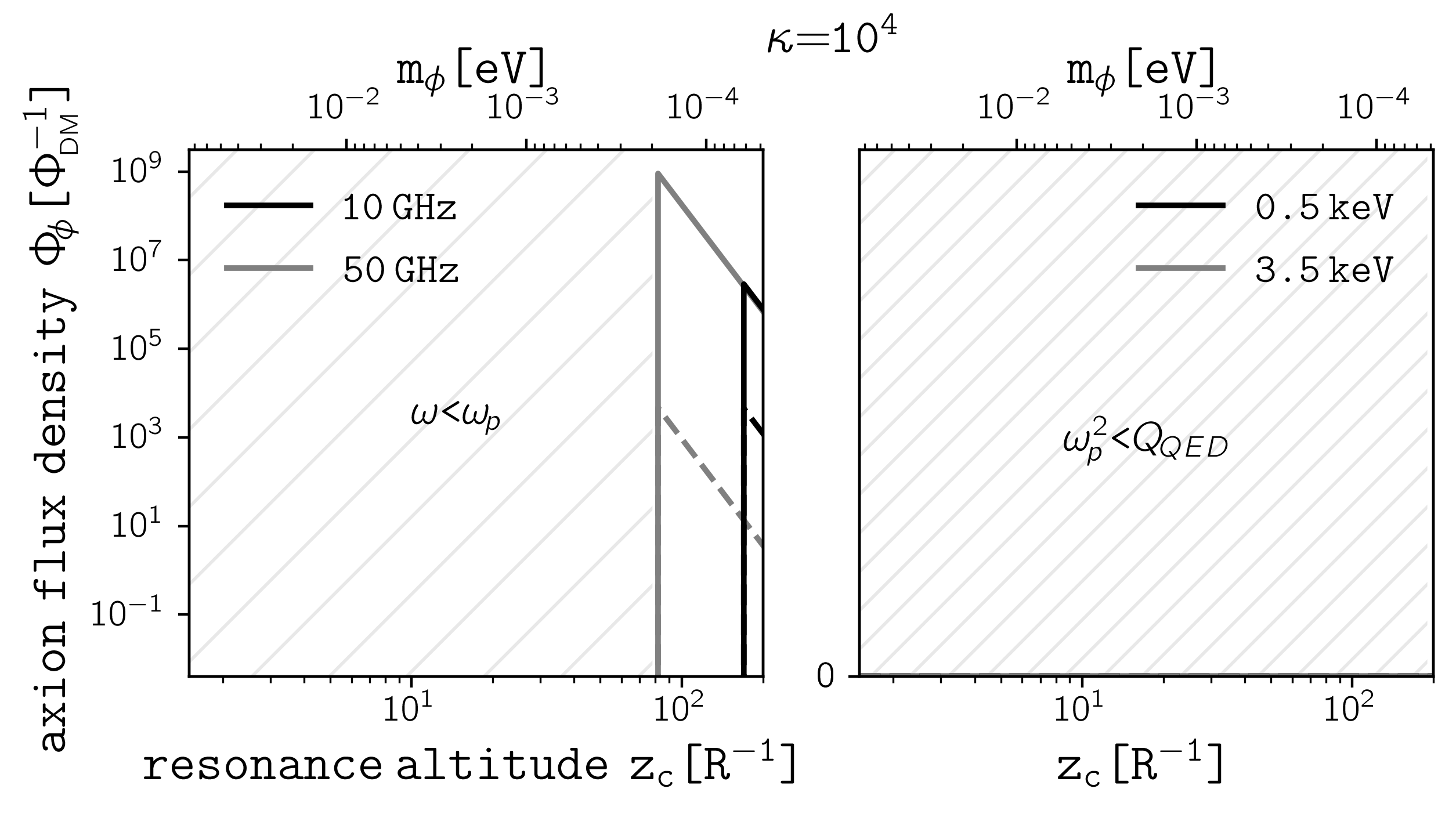}
    \label{fig:a}
\end{subfigure}
    \hfill
\begin{subfigure}{0.52\linewidth} \hspace{-12pt}
    \includegraphics[width=\linewidth, height=0.52\linewidth]{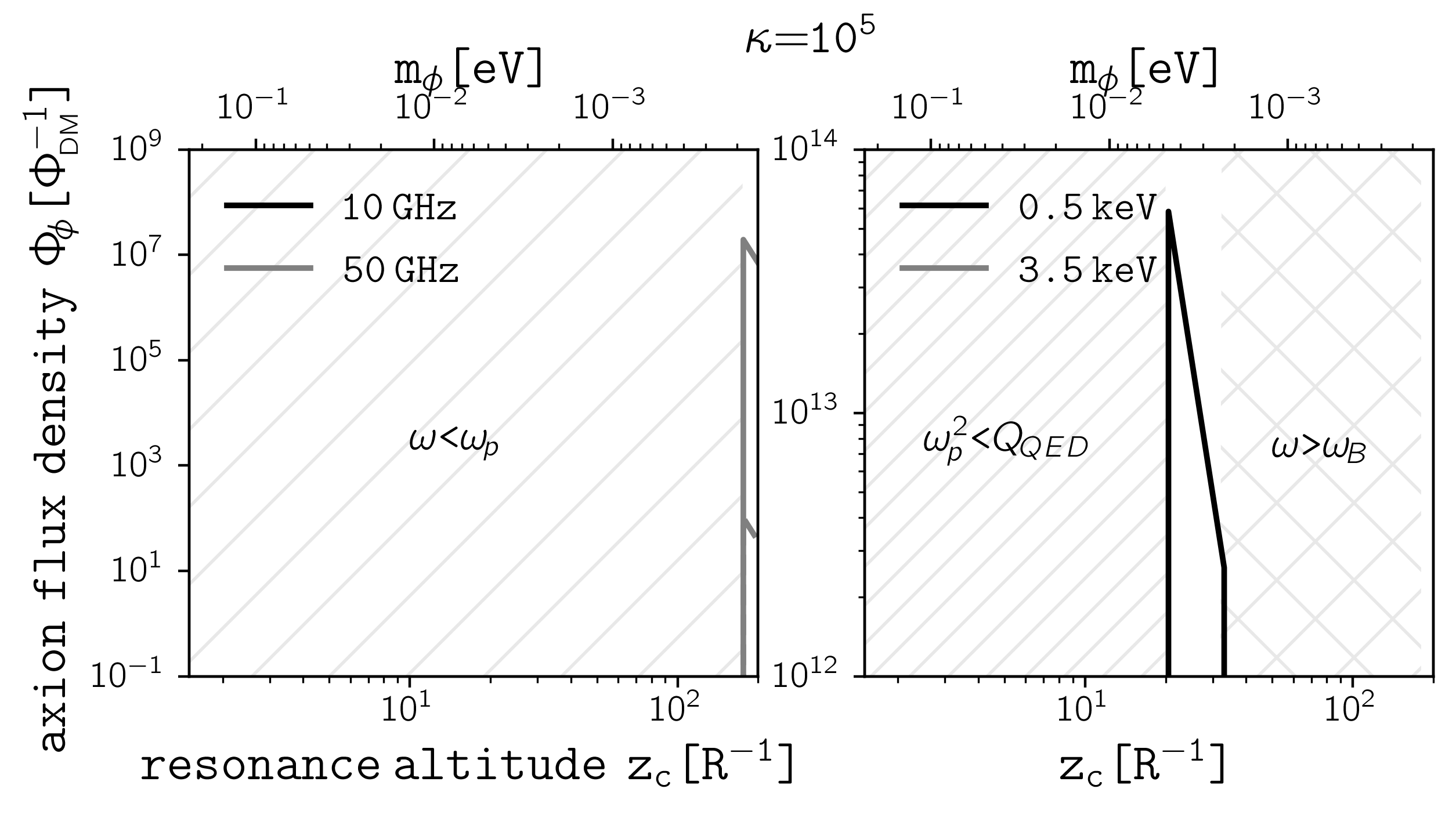}
    \label{fig:b}
\end{subfigure}

\begin{subfigure}{0.52\linewidth} \hspace{-10pt}
    \includegraphics[width=\linewidth, height=0.52\linewidth]{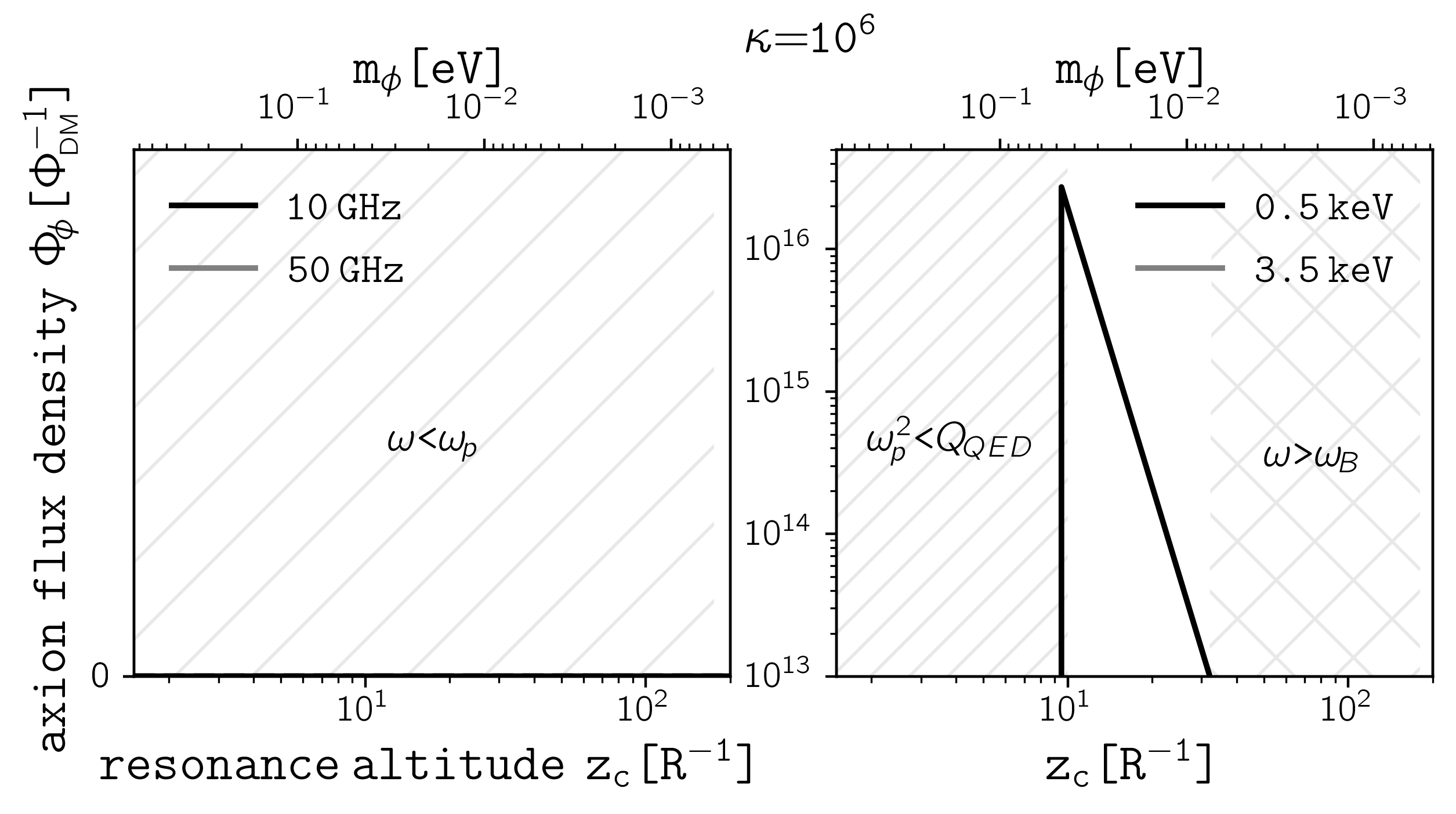}
    \label{fig:c}
\end{subfigure}
    \hfill
\begin{subfigure}{0.52\linewidth} \hspace{-12pt}
    \includegraphics[width=\linewidth, height=0.52\linewidth]{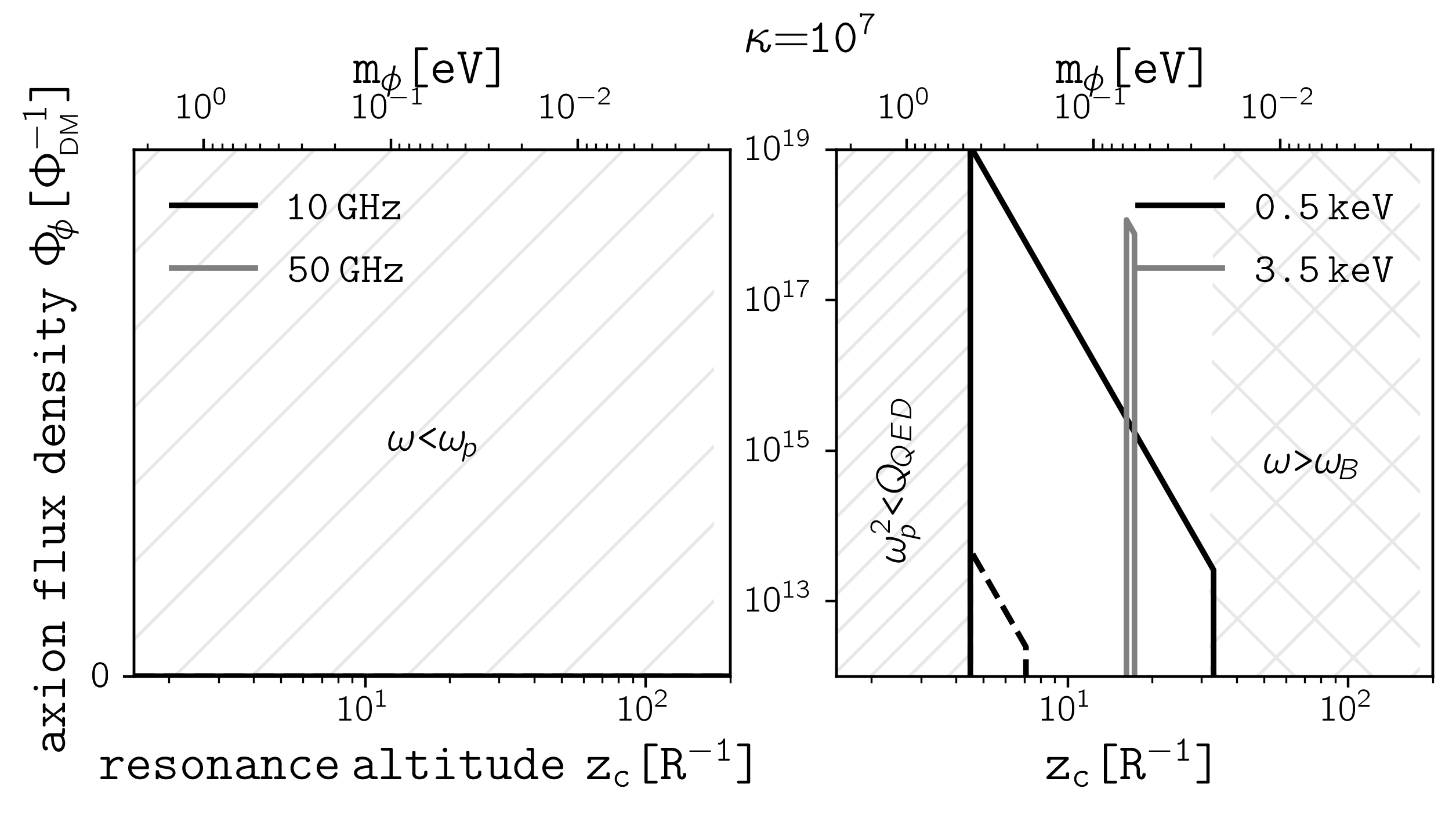}
    \label{fig:d}
\end{subfigure}
\caption{Quiescent axion flux through the surface of the light-cylinder of the star — $\Phi_{\!\phi}$ —, measured at the conversion point} in units of local dark matter density — $\Phi_{\!\mathrm{_{DM}}}$ — over resonance altitude — $z_c$ — or axion mass — $\mathrm{m}_{\phi}$ —. Starting at $h_{0}\!\gtrsim\! \nicefrac{R}{2}$ from the surface of the star \cite{Harding:1998ma} for several benchmark frequencies. The pair multiplicity factor varies between $\kappa\!=\!10^{4-7}$. Other parameters are $P$=1 s, $B_{\!_0}$ is $5\!\times\!10^8$ T for a rotation-powered pulsar — dashed line — or $B_{\!_0}\!=\!5\!\times\!10^{10}$ T for a magnetar — solid line —. We also adopt KSVZ-like coupling strength and canonical dimensions. The spectral densities were obtained from Fig. $\!\!$\ref{fig_2}. The resonant conversion in suppressed in sectors where O-modes become evanescent since $\omega_p\!>\!\omega$ overlapping adiabatic regions with $\Gamma \!\gtrsim\!1$; zones with $\omega\!\gtrsim\!\omega_{\!_B}$; the parameters space where QED vacuum effects govern the dispersion relations, or $\omega^2_p\!<\!Q_{\mathrm{QED}}$. 

\label{fig_4}
\end{figure*}

In Fig. $\!$\ref{fig_4} we report 1D calculations of the axion flux density induced by non-adiabatic resonant axion-to-photon conversion — $\Phi_{\!\phi}$ — in units of local dark matter flux density — $\Phi_{\!\mathrm{_{DM}}}\!$ — through the light cylinder of neutron stars for benchmark parameters. Several inferences can be drawn from the analysis of the different panels. For a fixed pair multiplicity factor, we consider both integrated radio-wave spectral density functions and persistent X-ray flux densities from Fig.$\,$\ref{fig_2}. The characteristic frequency of the plasma scales with the root of the charge density — see Eq.$\,$\ref{Eq.4.2} —. Consequently, near the stellar surface, for $\kappa\!\gtrsim\!10^6$ cut-off frequencies of the order of a few hundred GHz arise; and the corresponding parameter space is neglected for frequencies of a few tens of GHz as $\omega\!<\!\omega_p$ and hence O-modes are not propagated while the resonant conversion is disrupted — lower panels —. On the other hand, the ratio of plasma effects to vacuum effects on the mixing relations depends on both plasma density and photon energy — see appendix \ref{E} for details on the $\mathrm{QED}$ suppression of axion-photon cross-sections in magnetic stars —. From Eq.$\,$\ref{Eq.E.1}, it is found that axion-photon conversion is strongly suppressed below  $\kappa\!\!\lesssim\!\!10^{4-5}$ for keV photons since $\omega^2_p\!\!<\!<\!\!Q_{\mathrm{QED}}$ and vacuum birefringence governs the dispersion relations — upper panels —; or when $\kappa\!<\!10^7$ for photons with higher frequencies of several keV — bottom-right panel.

\section{Summary and conclusions}
\label{sec:Discussion}
In this article we have treated the non-adiabatic photon-to-axion resonant conversion in the state-of-the-art picture of neutron stars. Relying on observational data in both the radio-wave domain and the X-ray range, we have applied the model, of remarkable simplicity, to rotation-powered pulsars and magnetars. 

Within a realistic parameter space, results in Fig. $\!\!$\ref{fig_4} suggest that persistent flux densities up to $\lesssim$19 orders of magnitude over the dark matter flow in the Galactic halo, around 14--15 W$\,\!$m$^{-2}$, might originate somewhere in the polar caps of those astrophysical objects, region from which a dense beam of axions could be produced by photon-to-axion conversion. If we take into account that the relativistic axion velocity and the velocity dispersion of the surrounding dark matter differ by three orders of magnitude, this equates to dark matter overdensities of up to $\lesssim$16 orders of magnitude located somewhere in the vicinity of the star's light cylinder, in a large frequency-mass parameter space.

The conversion probabilities obtained from Eq.$\,$\ref{Eq.3.2} would be inaccurate in sectors in tension with the following limits: $\omega_p\!>\!\omega$, $\Gamma\!\gtrsim\!1$ and $\omega\!\gtrsim\!\omega_{\!_B}$ — i.e., sectors where O-modes are evanescent, adiabatic sectors or distant regions where the field becomes relatively weaker, respectively —. Consequently, we ruled out the model in these regions. In spite that the oscillation amplitudes tend to be lower beyond these limits and therefore our interest is reduced accordingly, methods that can be used to extend the model to those  sectors — shaded in Fig.$\,$\ref{fig_4} — have been suggested, e.g., multi-dimensional numerical calculations can extend the $\omega\!<\!<\!\omega_{\!_B}$ sector by enabling a `weak' field and also the propagation of oblique modes extending the $\omega_p\!>\!\omega$ space, ray-tracing technique, etc. \cite{Leroy:2019ghm, Witte:2021arp,Battye:2021xvt, Millar:2021gzs}. On the other hand,
the claim that Quantum Electrodynamics (QED) modifications to Maxwell’s equations due to vacuum polarization suppresses axion–(X-ray)photon conversion in highly magnetic neutron stars has been entrenched for many years \cite{Raffelt:1987im}. As a result of this, the possibility of relativistic axions with keV energies originating from photon-axion conversion in magnetic stars, or X-rays, bombarding the Earth causing observable effects has been neglected for decades. However, we have found, due to a higher pair density consistent with the current picture of neutron stars, a broad $\omega^2_p\!>\!Q_{\mathrm{QED}}$ sector at which the tension with QED vanishes in a more realistic neutron star model beyond the anachronistic Goldreich\&Julian’s density profile adopted firstly in \cite{Raffelt:1987im} and later in more contemporary works. The numerical approach of \cite{Fortin:2018ehg} would allow the simulations to be extended to the parameter space where vacuum polarization effects are relevant, i.e., $\omega^2_p\!<\!Q_{\mathrm{QED}}$, where the conversion probability is, however, suppressed by many orders of magnitude. Finally, some estimates might be in tension with the altitude from the stellar surface at which the radiation starts — i.e., regions with $z_c\!\lesssim\!h_{0}$ —. We note two points. First, different parametric configurations result in a higher/lower resonance altitude. Second, we have discussed in section \ref{sec:NS} that this parameter presents a significant uncertainty and can be ad hoc at times. Aiming at keeping our work as model-independent as possible, it was not ingrained more deeply throughout the manuscript.\newline

Neutron stars have the stronger magnetic fields known in the universe, about $B\!\lesssim \!10^{11}$ T at surface and axion convert efficiently to photon, and vice-versa, in a strong magnetic field. As ambient axions fall on the star, a narrow line, with $\Delta \nu \!<\!\!<\! \nu$, appears at a frequency corresponding to the axion mass. This signal has been searched for by different radio telescopes, without success. Unfortunately, the authors of the corresponding works adopted a star model with a Goldreich \& Julian (GJ) density profile to compare their observational data against a simulation that predicts the shape, frequency and intensity of the emission line originating from axion-to-photon conversion — see the most recent and restrictive results in \cite{Foster:2020pgt, Darling:2020uyo, Darling:2020plz} —. Crucially, astronomical observations of neutron stars and emission models allow one to estimate the observer-frame pair density, $n_e$, reliably and, by this method, the Goldreich \& Julian density profile adopted by those authors, dated in the late 1960s, has been shown to be under-estimated by many orders of magnitude in more recent works \cite{Guepin:2019fjb}. In other words, typical densities in neutron star magnetospheres concordant with the currently available data establish plasma densities above $\kappa \!\gtrsim\! 10^4$ times denser than the Goldreich \& Julian model, with $\kappa  \!=\!  n_e  /n_{_\mathrm{{GJ}}}$ the so-called ‘pair multiplicity factor’. This has important consequences. First, it is well-known that O-mode electromagnetic waves do not propagate in a plasma below a cut-off frequency, $\omega_p$, which scales with the plasma density profile with distance, $n_e(r)$, in the form $\omega_p\!\propto\! n_e^{\nicefrac{1}{2}}\!\!\rightarrow \omega_p\!\propto\! \kappa^{\nicefrac{1}{2}}$ — see Eq.$\,$\ref{Eq.4.2} —; hence suppressing the conversion of axions with a lower mass near the star surface, where the field would be stronger enhancing the conversion probability, P, as $\mathord{\mathrm{P}}\!\!_{\phi\shortto \gamma}\!\!\propto$B$^2$, since the conversion takes place at $m_{\phi}\!\sim\!\omega_p$ resonance. Moreover, such a large pair density implies that the exclusion bounds resulting from those previous works, set at $m_{\phi}\!\sim\!15\!-\!30$ $\mu$eV, are overestimated by several orders and, in fact, fall below the CAST exclusion sector and vanish \cite{CAST:2017uph}, since the resonance of axions of dozens of $\mu$eV of mass is only possible far away from the star surface, where the field is much weaker, as B$\propto\!\!r^{-3}$, and the conversion probability is reduced accordingly. Therefore, a significant blueshift of the spectral line emerging from axion-to-photon conversion is expected with respect to previous estimates, from frequencies of several GHz to frequencies of several hundred GHz, or $m_{\phi}\!\!\gtrsim\!\!100$ $\mu$eV. 
In consequence, caution should be exercised with earlier simulations that adopt the Goldreich \& Julian density profile in predicting radio signals originating from the resonant conversion of axions into O-mode photons in magnetospheres; while higher-frequency telescopes must be used for scanning, possibly leading to new exclusion limits for axion at high-frequency.\newline

The direct detection of neutron star axions at the Earth position by dark matter halo- helio- tele-scopes \cite{PhysRevLett.51.1415, PhysRevD.39.2089, DeMiguel:2020rpn} is an interesting question. Relativistic aberration transforms the regular semi-isotropic dipole pattern of radio-emission into a narrower beam of light with a high directionality in the observer's line-of-sight. The emerging axion beam is shaped accordingly because of momentum conservation, ideally. It also suffers from the inverse square law with distance. Geometric dilution would truncate prospects for the direct detection of the stellar axion flow even for features as giant radio pulses (GRPs) \cite{Jessner:2004ca,Karuppusamy:2010ar,Cordes:2003im} with a typical duration of the order of few-tens ns with a scatter time-scale of the order of the ms and a transient flux up to $10^{4-5}$ times over the unpulsed component observed by radio-telescopes. The high-energy frame — more specifically soft X-ray — should be examined in more detail. Dark matter astroparticles released from neutron stars could be analyzed in units of the axion rate received from the Sun \cite{PhysRevLett.51.1415}. The solar axion spectra presents a distribution peaking around 2-3 keV with a flux density about $10^{9}$ cm$^{-2}$s$^{-1}$ for $\mathrm{g}_{\phi\gamma}\!\sim 10^{-11}$ GeV$^{-1}$ \cite{Raffelt:1987np, PhysRevD.39.2089}. In this band, the quiescent fluence of magnetars can be $\pazocal{O}(10^{-1})$ cm$^{-2}$s$^{-1}$ — see Fig. $\!$\ref{fig_2} — while the photon-to-axion conversion amplitude can be $>\!\! 10^{-1}$ for a broad parameter space. Thus, solar axion flux would be $\lesssim\!\!10^{12}$ times denser than axion magnetar flux at the Earth position due to geometric dilution. Nonetheless, intermediate bursts (IBs) featured by magnetars, with luminosities up to $\gtrsim \!\!10^{43}$ erg$\,$s$^{-1}$ and a characteristic time of tens of seconds, and mostly giant flares (GFs), which present a spike lasting $\sim$0.1--1 s that can reach powers up to $\gtrsim \!\!10^{47}$ erg$\,$s$^{-1}$ and a long tail — few hundred seconds — have been observed at $\hbar\omega\!\!>\!\!2$ keV \cite{Turolla:2015mwa,Kaspi:2017fwg,Mereghetti:2009haa, Mereghetti:2005dt,Hurley:2005zs,Boggs:2006uk,Roberts:2021udn}. They represent a contrast to their quiescent luminosities in the range 2-10 keV, with a break about $\lesssim \!\! 10^{33}$ erg$\,$s$^{-1}$ for  — the less bright — 
`transient' magnetars, and $\gtrsim \!\! 10^{33}$ erg$\,$s$^{-1}$ for — the more potent — `persistent' sources. In spite of we emphasize that this result must be viewed with caution,\footnote{Note that IBs and GFs could involve different physics than the main considerations in this paper.} numbers could add up so that the direct detection of magnetar axions triggered by X-ray IBs/GFs cannot be discarded \cite{DeMiguel:2022ojb}. \newline

\section*{Acknowledgements}
JDM would like to thank J. Eilek and R. Génova for comments about observational aspects of radio-astronomy. This work was supported by the Special Postdoctoral Researchers (SPDR) program.

\appendix
\section{Synchrotron self-absorption spectra of optically thin sources}
\label{A}
The empirical spectra of radio-pulsars can be fitted with a power law in the form $\pazocal{S}_{\!\nu}\!\sim\! \nu^{- \alpha}$. The `spectral index' $\alpha$ adopts a typical value of the order of $\alpha \!\!\lesssim$1 at frequencies $\nu\!\!\lesssim\!\!\nu_b$, and $<\!\!\alpha\!\!>\sim\!\!1.8$ for $\nu\!\!\gtrsim\!\!\nu_b$, where $\nu_b$ is commonly referred to as the `break' frequency. The break frequency is typically $\nu_b \!\lesssim\!$ 500 MHz, although it presents certain variability, extending to higher frequencies in some cases \cite{Maron:2000wn,Jankowski:2017yje,Bates:2013ear,Murphy:2017ech}. Plasma restrain the synchrotron emission, forcing the spectral distribution to curve down at low frequency \cite{1970ranp.book.....P}. As a result, the low-frequency self-absorption spectral function of an
optically thick pulsar presents a power-law with a very steep slope 

\begin{equation}
\pazocal{S}_{\!\nu}= \frac{2}{3}\! \,\left(\frac{\nu}{\nu_b}\right)^{\!\!5/2}\left[1-\mathrm{exp}\Bigg\{\!\!\left(-\frac{\nu}{\nu_b}\right)^{\!\!-(\delta+4)/2}\Bigg\} \right]\,.
\label{Eq.2.1}
\end{equation}

The spectral indexes are related by $2\alpha+1=\delta$. For simplicity, we have approximated $\omega_{b}\!\sim\! \omega_{\tau}$, corresponding $\omega_{\tau}$ to the frequency which maximizes the optical path. This assumption would lead to deviations within the precision that is desired for the present work.

\section{Dielectric tensor of a cold plasma in a strong homogeneous magnetic field}
\label{C}
Longitudinal modes in a plasma are compressive waves. In the equation of motion, their respective restoring force is described by a $\nabla$p term. For considering a cold homogeneous plasma, we neglect the $\nabla$p pressure. With $\hat{\mathrm{m}}\!\!\equiv\!\!\hat{\mathrm{z}}$ the coordinate along the magnetic field $B$, $\mathrm{\hat{x}}$ and $\mathrm{\hat{y}}$ directions across the magnetic field, the effective dielectric tensor of a cold plasma presents the form \cite{1993ppm..book.....B}

\begin{subequations}
\begin{align}
\label{Eq.c1}
\varepsilon_{\mathrm{xx}}&=\varepsilon_{\mathrm{yy}}=1+\left<\frac{\omega^2_p\gamma\tilde{\omega}^2}{\omega^2(\omega^2_{\!_B}-\gamma^2 \tilde{\omega}^2)} \right> ,\\
\label{Eq.c2}
\varepsilon_{\mathrm{xy}}&=-\varepsilon_{\mathrm{yx}}=i\left<\frac{\omega^2_p\omega_{\!_B}\tilde{\omega}}{\omega^2(\omega^2_{\!_B}-\gamma^2 \tilde{\omega}^2)} \right>  \, ,\\
\label{Eq.c3}
\varepsilon_{\mathrm{xz}}&=-\varepsilon_{\mathrm{zx}}=\left<\frac{\omega^2_p\gamma k_x v_{_{\!\parallel}}\tilde{\omega}}{\omega^2(\omega^2_{\!_B}-\gamma^2 \tilde{\omega}^2)} \right>  \, ,\\
\label{Eq.c4}
\varepsilon_{\mathrm{yz}}&=-\varepsilon_{\mathrm{zy}}=-i\left<\frac{\omega^2_p \omega_{\!_B} k_x v_{_{\!\parallel}}}{\omega^2(\omega^2_{\!_B}-\gamma^2 \tilde{\omega}^2)} \right>  \, ,\\
\label{Eq.c5}
\varepsilon_{\mathrm{zz}}&=1-\left<\frac{\omega^2_p}{\gamma^3\tilde{\omega}^2}\right>+\left<\frac{\omega^2_p \gamma k^2_x v^2_{_{\!\parallel}}}{\omega^2(\omega^2_{\!_B}-\gamma^2 \tilde{\omega}^2)} \right>    \,;
\end{align}
\end{subequations}

where $\omega_p$ is defined by Eq. $\!$\ref{Eq.4.2}, $\tilde{\omega}=\omega-k_z v_{_{\!_{\parallel}}}$, $\gamma=(1-v^2_{\!_{\parallel}}/c^2)^{\nicefrac{-1}{2}}$ and $\omega_{\!_B}\!=\!e\!\!\mid \!\!B\!\! \mid\!\!/m_e$ with $\mid \!\!B\!\! \mid =\! B_{\!_0}$, $e$ the elementary charge and $m_e$ the electron mass.

The electrodynamic properties of a pulsar magnetosphere can be described in the strong field limit $B_{\!_0}\!\!\rightarrow\!\infty$, where $\omega_{\!_B}\!\!>\!>\!\omega$ and $\omega_{\!_B}\!\!>\!>\!\omega_p$. In this approximation, easily fulfilled in the vicinity of the star, where the field is stronger, or for sufficiently low frequency, the dielectric permittivity becomes $\varepsilon_{\mathrm{xx}}\!=\!\varepsilon_{\mathrm{yy}}\!=\!1$, $\varepsilon_{\mathrm{xy}}\!=\!\varepsilon_{\mathrm{yx}}\!=\!\varepsilon_{\mathrm{yz}}\!=\!\varepsilon_{\mathrm{zy}}\!=\!0$, and 

\begin{subequations}
\begin{align}
\varepsilon_{\mathrm{zz}}&=1-\left<\frac{\omega^2_p}{\gamma^3\tilde{\omega}^2}\right>\,.
\end{align}
\end{subequations}

This ensures that the conductivity is non-zero only along the direction of the magnetic field.

\section{The role of the density profile in the weight of QED corrections }
\label{E}
The claim that QED modifications to Maxwell's equations due to vacuum polarization suppresses axion--X-ray conversion in highly magnetic stars has been entrenched for decades \cite{Raffelt:1987im}. However, the tension with QED is relaxed in the modern neutron star model beyond the Goldreich \& Julian density profile adopted throughout this manuscript \cite{1969ApJ...157..869G, Guepin:2019fjb}. 

Corrections due to QED effects arise from the Euler-Heisenberg Lagrangian $\pazocal{L}_{\mathrm{QED}}=\frac{\alpha^{\!2}}{90m^{\!4}_{\!e}}[\left(F_{\!\mu\nu}F^{\mu\nu}\right)^{2}+\frac{7}{4}(F_{\!\mu\nu}\tilde{F}^{\mu\nu})^{2}]$ in strong external fields. Thus, the leading order terms of the refractive indices for plasma and vacuum are $\Delta_p\!\sim\!\nicefrac{\omega^2_p}{2\omega}$ and $\Delta_{\mathrm{{vac}}}\!\sim\!  \frac{7\alpha}{90\pi}\omega\!\left (\!\nicefrac{B(r)}{B_{\mathrm{crit}}}\right)^2$, respectively; being $\alpha\!=\! \nicefrac{e^2}{4\pi}$ the fine-structure constant and $B_{\mathrm{crit}}\!=\!\nicefrac{m^2_e}{e}$ the `critical' field. Therefore, the ratio of plasma effects to vacuum effects — $\nicefrac{\Delta_p}{\Delta_{\mathrm{{vac}}}}$ — is read \cite{Huang:2018lxq}

\begin{equation}
\frac{\omega_{p}^2}{Q_{\mathrm{{QED}}}} = 5\!\times\!10^{8}\left(\!\!\frac{\mathrm{\mu eV}}{\omega}\!\right)^{\!\!2}\frac{\mathrm{10^{8} [T]}}{ B(r)}\frac{\mathrm{1[s]}}{P}\,\kappa \, ,
\label{Eq.E.1}
\end{equation}

where we have introduced the plasma frequency profile from Eq.$\,$\ref{Eq.4.2} and defined $Q_{\mathrm{{QED}}}=2\omega\Delta_{\mathrm{{vac}}}$. From the analysis of Eq.$\,$\ref{Eq.E.1} it follows that no observable vacuum birefringence effects are expected for low-frequency photons, as $\omega^2_p>\!>Q_{\mathrm{QED}}$ and plasma effects govern the dispersion relations in Eq.$\,$\ref{Eq.3.1}; while in the case of high-energy photons the weight of vacuum polarization effects gradually vanishes at relatively large distances from the stellar surface for pair multiplicity factors above $\kappa\!\gtrsim\!\!10^5$, consistent with the current picture of neutron stars.

\section{Axion electrodynamics in neutron stars}

A modification of Maxwell's equations arises from a light, pseudo-stable axion \cite{Wilczek}

\begin{subequations}
\begin{align}
\nabla \cdot (\mathbf {E} +\mathrm{g}_{\phi\gamma} \phi \mathbf {B} )&=\rho \, ,\\
\nabla \times (\mathbf {B} -\mathrm{g}_{\phi\gamma} \phi \mathbf {E} )&=\dot{\mathbf {E}} +\mathrm{g}_{\phi\gamma} \dot{\phi} \mathbf {B} +\mathbf {J} \, ,\\
\nabla \cdot \mathbf {B}&=0 \, ,\\
\nabla \times \mathbf {E} + \dot{\mathbf {B}}  &= 0 \,;
\end{align}
\end{subequations}
where the monopole density $\rho_m$ and the monopole current $\mathbf{J}_{\!m}$ were replaced by zero. 

From the linearisation of the equations of motion from Eq. $\!$\ref{Eq.1.1} for a stationary and free of electric charge neutron star background in the aligned rotator approximation, the following system of equations emerges \cite{Battye:2019aco}
\begin{subequations}
\begin{align}
\label{Eq.3.2a}
\left(\Box +\mathrm{m}_{{\phi}}^{2}\right)\!\phi &= \mathrm{g}_{\phi\gamma} \,\mathbf {E} \cdot \mathbf {B}_0 \, , \\
\Box \mathbf {E}+ \nabla (\nabla\cdot\mathbf {E})+ \mathbf{\sigma}\dot{\mathbf{E}} &= - \mathrm{g}_{\phi\gamma} \ddot{\phi}\mathbf {B}_0 \, ,
\label{Eq.3.2b}
\end{align}
\end{subequations}

where it was introduced $\mathbf {A} \!\leftarrow \!\mathbf {A}_0+\mathbf {A}$ in fields and densities while $\mathbf{J}=\mathbf{\sigma}\cdot \mathbf{E}$, being $\mathbf{\sigma}$ the conductivity tensor. Note the Klein-Gordon formalism in Eq. $\!$\ref{Eq.3.2a}.

\section{Gradient of the plasma frequency in the magnetosphere of neutron stars}
\label{D}
In the aligned rotator limit — $\hat{\mathrm{m}}\simeq\hat{\mathrm{z}}$ — and on-the-spot resonance approximation — where the magnetic field presents a negligible gradient —, from Eqs. \ref{Eq.4.1} and \ref{Eq.4.2} we get

\begin{equation}
\omega_p\simeq \overbracket{\left( \frac{2 e\Omega B_{\!_0} \kappa}{m_e} \right)^{\!\!\!1/2}}^{\xi} z^{-3/2} .
\label{Eq.d.1}
\end{equation}

In that form, it follows that $\frac{\partial}{\partial z}\omega_p = -\nicefrac{3}{2} \,z^{-5/2} \xi$. At resonance — $\omega_p\!=\!\mathrm{m_{\phi}}$ — it is straightforward to obtain

\begin{equation}
\frac{\partial\omega_p}{\partial z} = -\frac{3}{2} \frac{\mathrm{m_{\phi}}}{z_c} \, ,
\label{Eq.d.2}
\end{equation}

where we have used $\mathrm{m_{\phi}}=z^{-3/2}_c\xi $ from Eq. \ref{Eq.d.1}.




\end{document}